\documentclass[a4paper,12pt]{article}

\usepackage{graphics,amsmath,amssymb,epsfig,epstopdf}

\begin{document}

\title{
\bf  Transient dark energy, cosmological constant boundary crossing and dark energy $w(z)$ data fits}

\author{Dalibor Perkovi\'{c}$^{1,}$\thanks{dalibor.perkovic@zvu.hr}  and Hrvoje \v Stefan\v ci\'c$^{2,}$\thanks{hrvoje.stefancic@unicath.hr}}


\vspace{3 cm}
\date{
\centering
$^{1}$ University of Applied Health Sciences, Mlinarska street 38, 10000 Zagreb, Croatia \\
\vspace{0.2cm}
$^{2}$ Catholic University of Croatia, Ilica 242, 10000 Zagreb, Croatia }


\maketitle

\abstract{ The formalism of dark energy based on modeling speed of sound as a function of equation of state parameter is elaborated. A specific model which allows detailed study of cosmological constant boundary crossing is introduced and analytical solutions for the model dynamics are obtained. It is shown how in specific parameter regimes dark energy can be a transient phenomenon. It is further demonstrated how the model reproduces specific features of recent fits of dark energy $w(z)$ to observational data.}

\vspace{2cm}

\section{Introduction}

\label{intro}

The dynamical nature of the observable universe is primarily reflected in its global expansion. Viable scenarios of cosmic expansion history consistent with numerous available datasets \cite{SNIa1,SNIa2,CMB1,CMB2,BAO} include a sequence of epochs of decelerated and accelerated expansion. As decelerated expansion follows naturally from components such as radiation and nonrelativistic matter (dust), familiar from terrestrial experiments and observations, efforts to understand mechanisms of accelerated cosmic expansion have attracted much more attention of cosmologists in the past decades.  The epoch of inflation \cite{Inflation1,Inflation2,Inflation3}, assumed to set up the initial seeds of structures in our observable universe, is closely related to the accelerated expansion of the universe. The observational data also indicate that the late-time epoch of accelerated expansion started at a redshift $z < 1$ and is still ongoing \cite{SNIa1,SNIa2}. 

Whereas the accelerated character of expansion in the present stage of cosmic evolution seems well established, the mechanism behind the acceleration is far from being equivocally determined. In a multitude of ingenious approaches and models two groups stand out: (i) dark energy models which assume the existence of a cosmic component with a sufficiently negative pressure \cite{DERev1,DERev2,DERev3,DERev4,DERev5} and (ii) modified gravity models which propose that gravitational interaction is modified at cosmic distances \cite{ModGravRev1,ModGravRev2,ModGravRev3}. In many models from both groups important challenges are posed by various forms of singularities \cite{Tsujikawa}. Singular phenomena that have attracted a lot of attention comprise Big Rip \cite{cald1,cald2}, sudden future singularities \cite{Barrow, Stef2004}, Little Rip \cite{LR1,LR2,LR3} and Pseudo-Rip \cite{PR} among others.   

A promient question related to epochs of accelerated expansion is their end. In the case of inflation, a so called graceful exit from inflatory expansion is required to allow the evolution of inhomogeneities created in the last e-folds of inflatory expansion. In majority of dark energy models, the dark energy density begins dominating at a redshift $z < 1$ and it remains dominant throughout the remaining expansion of the universe. This may lead either to singularities (as in the case of  phantom energy with constant parameter of equation of state (EoS)), or to formation of future horizons. Therefore, it is of considerable interest to learn if the present late-time epoch of acceleration is transient or the expansion remains accelerated throughout the remaining cosmic expansion. The phenomenon of transient cosmic acceleration was studied in numerous works. For some interesting models of transient acceleration see \cite{Bean,Russo,Fuzfa,Alcaniz,Koivisto,Odintsovtransition}.

In the course of cosmic dynamics, cosmic components/fluids interchange in the role of the dominant component (which dictates the type of the expansion of the universe) owing to their different equation of state, which leads to different scaling with the expansion. In this way the status of dominant component shifts with expansion towards components with smaller values of the parameter of EoS and thus results in a sequence of radiation dominated, nonrelativistic matter dominated and dark energy dominated phases. In such a setting cosmic components retain their type of scaling of energy density with the scale factor - decreasing or increasing. In this paper we are interested in a different sort of scaling of component energy density. In particular, we are interested in a $\rho_d(a)$ function which first increases with the expansion up to some specific value of the scale factor and then decreases. We call this type of component a transient component. As we are primarily interested in dark energy dynamics, in the remainder of this paper we shall refer to this component as {\bf transient dark energy}. Nevertheless, the concept of a transient component is clearly also of interest for other accelerated epochs of cosmic expansion such as epoch of inflation.   A recent attempt by Park \cite{Park}, called DE-spike model, aimed at identifying better fits to observational data, includes an example of transient dark energy where dark energy exhibits a spike at a redshift close to radiation-matter energy density equality. 

It is also important to stress that concepts of transient cosmic acceleration and transient dark energy, though closely related, are not identical. In particular, in principle it is possible to have transient cosmic acceleration where the dark energy density is strictly descending function. An example would be a dark energy component with $-1 < w < -1/3$ up to now, causing the present accelerated cosmic expansion, which transits to $w>-1/3$ at some moment in the future. Vice versa, transient dark energy need not neccessarily lead to transient acceleration. An example of such a situation would be a dark energy model smoothly interpolating between $w_1<-1$ in the distant past and $-1<w_2<-1/3$ in the distant future.   

In this paper we focus on studying the equation of state of a single cosmic component exhibiting the said transient behavior. In the remainder of the paper we omit the subscript $d$ in quantities describing the transient dark energy since we do not consider other cosmic components.  The evolution of energy density with the expansion
\begin{equation}
\label{eq:conserv}
\frac{d \rho}{\rho} = -3 (1+w) \frac{ d a}{a} \, ,
\end{equation}
reveals that the extremum of the energy density evolution can be achieved at $w=-1$. Moreover, the required sequence of energy density first growing, then reaching some maximal value and then  decreasing can be realized if the EoS parameter evolves from values below -1, crosses the cosmological constant (CC) boundary and then proceeds towards values above -1. This indicates that a single component transient dark energy inevitably requires the CC boundary crossing from the phantom-like to the quintessence-like region.  Therefore, a suitable framework for the study of transient single fluid dark energy also needs to incorporate the CC boundary crossing as a generic phenomenon, as also mentioned in \cite{Park}.

The phenomenon of CC boundary crossing has been studied in various frameworks \cite{PRD2005,quintom1,quintom2,kess,saridakis}, for a review see \cite{saridakisrev}. The analysis of the CC boundary crossing in the effective single field dark energy model was made in \cite{creminelli}. The realization of CC boundary crossing in imperfect fluids was addressed in \cite{vikman2} and the (divergent) behavior of quantum perturbations at the CC boundary crossing was analyzed in \cite{timecrystal1}.
The effect of CC boundary crossing has also recently attracted considerable attention as it is a necessary ingredient of cosmic time crystal models \cite{wilczek,timecrystal1,timecrystal2}.  The particular phenomenon of CC boundary crossing for a single cosmic component was previously studied in \cite{PRD2005} where it was shown that the required equation of state needs to be implicitly defined. Although a regular modeling framework, the cosmic component implicit equation of state is not always practical for modeling of the evolution of $w$ with the expansion. Therefore, instead of following the approach introduced in \cite{PRD2005}, we adopt a more suitable approach of modeling of the square of the cosmic component speed of sound as a function of the EoS parameter, i.e. $c_s^2=c_s^2(w)$. This approach has so far been successfully applied to introduction and analysis of generalized models of unification of dark matter and dark energy in \cite{PRD2013}.

Our physical motivation for considering models of the type $c_s^2=c_s^2(w)$ is threefold. In particular, we would like
\begin{itemize}
\item to study the phenomenon of cosmological constant boundary crossing in a more detail to gain deeper insight into properties of this potentially very important part of global cosmological dynamics;
\item to construct explicit models of transient cosmic component in the sense explained above;
\item to consider a possible explanation of a recent fit for dark energy model with a general redshift dependence of the EoS parameter, $w(z)$, presented in \cite{NatureAstronomy}.   In subsection \ref{disc2} it will be shown how the model studied in this paper reproduces general features of $w(z)$ obtained in \cite{NatureAstronomy} in a specific parameter regime.    
\end{itemize}

 In this paper we expand the analysis initiated in \cite{PRD2013} and examine the CC boundary crossing  in the framework of a single cosmic component with a general EoS defined by 
 $c_s^2=c_s^2(w)$. When thinking of this cosmic component, it is natural to consider the description in terms of a barotropic cosmic fluid since
\begin{itemize}
\item barotropic cosmic fluid  is the simplest description of a cosmic component which is used for the description of the ``standard'' components such as nonrelativistic matter and radiation. It is important to assess suitability of the barotropic fluid framework for the CC boundary crossing modeling in our approach and to establish its strengths and weaknesses;
\item barotropic cosmic fluid framework can be considered as an effective description of the more elaborate microscopic dynamics.  Analytic solutions obtained in the barotropic fluid framework can then be used to reconstruct underlying fundamental models. Some calculations in this direction are presented in section \ref{scalar}.
\end{itemize} 

The paper has the following organization. After the introduction given in section \ref{intro}, section \ref{model} brings the presentation of the dark energy model and the solutions of its dynamics in terms of scale factor. In section \ref{scalefactor} the closed form solution for the dynamics of the scale factor with cosmic time is obtained and the appearance of finite-time singularities is studied. Section \ref{scalar} is focused on scalar field representation of the model dynamics in quintessence-like and phantom-like regimes. In section \ref{disc} various parameter regimes are discussed and the potential of the model in explaining some of recent fits of dark energy redshift dependent equation of state is outlined. The paper closes with conclusions given in section \ref{concl}.

\section{The model}

\label{model}

We adopt a formalism introduced in \cite{PRD2013} based on modeling the cosmic component speed of sound in terms of its EoS parameter, i.e.

\begin{equation}
\label{eq:cs2}
c_s^2=\frac{d p}{d \rho} = c_s^2 (w) \, .
\end{equation}

An alternative view on this approach in modeling the dark energy EoS, which does not rely on the concept of speed of sound, is that (\ref{eq:cs2}) expresses $d p/d \rho$ as a function of $p/\rho$. Such a view is more appropriate when the description based on (\ref{eq:cs2}) is considered as an effective approach which facilitates model analysis, in particular finding analytic solutions.

Using the fact that $p=w\rho$, Eq.  (\ref{eq:cs2}) yields

\begin{equation}
\label{eq:rhodyn}
\frac{d \rho}{\rho}=\frac{d w}{c_s^2 (w) - w} \, .
\end{equation}

Inserting (\ref{eq:rhodyn}) into (\ref{eq:conserv}) results in

\begin{equation}
\label{eq:cs2ofw}
\frac{d w}{(c_s^2 (w)-w)(1+w)} = - 3 \frac{d a}{a} \, .
\end{equation}

This paper is centered on the following model:

\begin{equation}
\label{eq:model}
c_s^2 (w)=w + A (1+w)^B \, .
\end{equation}

This choice represents an extension of the barotropic fluid model with a constant EoS. Namely, for the equation of state $p=w\rho$, with $w$ constant, we have $c_s^2=w$. The model (\ref{eq:model}) therefore represents a generalized correction to the constant EoS case. With a sufficently small $A$, the model (\ref{eq:model}) can be made indistinguishable from the constant EoS case.  

In our considerations of the model (\ref{eq:model}) we restrict our analysis of the values of the parameter $B$ to the ratio of two odd numbers, i.e. $B=\frac{2 n +1}{2 m+1}$. This choice ensures that all relevant functions (such as $w(a)$, $\rho(a)$, etc.) are well defined for the entire interval of the scale factor. On the other hand, any real value of the parameter $B$ can be arbitrarily precisely approximated by a ratio of two odd numbers and in this way we may be confident that no specific dynamical regime dictated by $B$ might be missed by our practical restriction on $B$.

For the model (\ref{eq:model}), it is also useful to write the evolution equation for $w(a)$ as

\begin{equation}
\label{eq:wofaexplicit}
a \frac{d \, w}{d \, a} = -3 A (1+w)^{B+1} \, .
\end{equation}

For the parameter values used in this paper, $B=\frac{2 m+1}{2 n + 1}$, the exponent in (\ref{eq:wofaexplicit}) has the value $B+1=\frac{2(m+n+1)}{2 n +1}$. This fact reveals that the derivative in the l.h.s of (\ref{eq:wofaexplicit}) is of the same sign, wherever it is defined, even at the CC boundary crossing point where $w+1$ changes sign.

For the nontrivial values $A \neq 0$, $B \neq 0$ and $B \neq 1$, inserting (\ref{eq:model})  into (\ref{eq:cs2ofw}) leads to
\begin{equation}
\label{eq:wofa}
w = -1 + \left[ (1+w_0)^{-B} + 3 A B \ln \frac{a}{a_0} \right]^{-\frac{1}{B}} \, .
\end{equation}

On the other hand, Eq. (\ref{eq:rhodyn}) can be readily integrated to obtain an explicit function $\rho(w)$, i.e.

\begin{equation}
\label{eq:rhoofw}
\rho=\rho_0 \; e^{\frac{(1+w)^{-B+1}-(1+w_0)^{-B+1}}{A(-B+1)} } \, .
\end{equation}

The expression (\ref{eq:wofa}) can be inserted in (\ref{eq:rhoofw}) to yield an explicit expression for energy density as a function of the scale factor

\begin{equation}
\label{eq:rhoofa}
\rho=\rho_0 \; e^{ \frac{1}{A(1-B)} \left( \left[ (1+w_0)^{-B} + 3 A B \ln \frac{a}{a_0} \right]^{\frac{B-1}{B}} - (1+w_0)^{-B+1} \right)} \, .
\end{equation}

As already indicated, solutions (\ref{eq:rhoofw}) and (\ref{eq:rhoofa}) are not directly applicable for special values $B=0$ and $B=1$. Since $B=0$ is not of the form $B=\frac{2 m+1}{2 n + 1}$ used in this paper, its further discussion is not needed. For the value $B=1$, the solution for $w(a)$ given in (\ref{eq:wofa}) is directly applicable, whereas the relation $\rho(w)$ acquires the form
\begin{equation}
\label{eq:rhoofwB1}
\rho=\rho_0 \left| \frac{1+w}{1+w_0} \right|^{\frac{1}{A}} \, .
\end{equation}

Finally, the result (\ref{eq:wofa}) can be inserted into (\ref{eq:model}) to obtain the scaling of the sound speed squared
\begin{equation}
\label{eq:cs2ofa}
c_s^2=-1 + \left[ (1+w_0)^{-B} + 3 A B \ln \frac{a}{a_0} \right]^{-\frac{1}{B}} + A \left[ (1+w_0)^{-B} + 3 A B \ln \frac{a}{a_0} \right]^{-1} \, .
\end{equation}

\section{Scale factor dynamics}

\label{scalefactor}

As our model defined in (\ref{eq:model}) allows an explicit solution for $\rho(a)$ given in (\ref{eq:rhoofa}), in situations where the dark energy component dominates the expansion of the universe (i.e. the contribution of other cosmic components is negligible), Friedmann equation (we perform our analysis for the spatially flat universe, $k=0$)
\begin{equation}
\label{eq:Friedmann}
H^2=\frac{8\pi G}{3} \rho \, ,
\end{equation}
can be writen as
\begin{equation}
\label{eq:Friedexplicit}
\frac{d}{d t} \left( \ln \frac{a}{a_0} \right) = \sqrt{\frac{8 \pi G}{3} \rho_0} e^{\frac{1}{2 A (1-B)} \left[ \left[(1+w_0)^{-B}+3 A B \ln \frac{a}{a_0} \right]^{\frac{B-1}{B}} - (1+w_0)^{1-B} \right] } \, .
\end{equation}
This equation acquires a somewhat more tractable form in terms of a new variable $y=(1+w_0)^{-B}+3 A B \ln \frac{a}{a_0}$, in particular
\begin{equation}
\label{eq:Fried_y}
\frac{d y}{d t}=3 A B  \sqrt{\frac{8 \pi G}{3} \rho_0} e^{-\frac{(1+w_0)^{1-B}}{2 A (1-B)}}  e^{\frac{1}{2 A (1-B)} y^{\frac{B-1}{B}}} \, .
\end{equation}
This equation can be integrated to give a closed form expression for expression $t=t(a)$:
\begin{equation}
\label{eq:tofa}
t-\tilde{t}=\frac{1}{3 A B} \sqrt{\frac{3}{8 \pi G \rho_0}}  e^{\frac{(1+w_0)^{1-B}}{2 A (1-B)}} \int_{(1+w_0)^{-B}+3 A B \ln \frac{\tilde{a}}{a_0}}^{(1+w_0)^{-B}+3 A B \ln \frac{a}{a_0}} e^{-\frac{1}{2 A (1-B)} y^{\frac{B-1}{B}}} \, d y \, ,
\end{equation} 
where the dark energy component (\ref{eq:rhoofa}) dominates the expansion in interval $(\tilde{a},a)$ and $\tilde{t}$ is the cosmic time corresponding to the scale factor $\tilde{a}$. 

We are in particular interested in the value of $t$ when the scale factor tends to infinity i.e. $t_{\infty}=\lim_{a \rightarrow \infty} t$. For simplicity let us take $\tilde{a}=a_*$, where $(1+w_0)^{-B}+3 A B \ln \frac{a_*}{a_0} = 0$. Using the fact that for selected values of parameter $B$ we have $(-1)^{\frac{B-1}{B}}=1$, it is possible to write (\ref{eq:tofa}) for all values of parameters $A$ and $B$ as
\begin{equation}
\label{eq:tinfty}
t_{\infty}-t_*=\frac{1}{3 | A B|} \sqrt{\frac{3}{8 \pi G \rho_0}}  e^{\frac{(1+w_0)^{1-B}}{2 A (1-B)}} \int_0^{\infty} e^{-\frac{1}{2 A (1-B)} y^{\frac{B-1}{B}}} \, d y \, .
\end{equation}
The analysis of this expression yields the following results:
\begin{enumerate}
\item for $A(1-B) < 0$ (which corresponds to cases $A>0, B>1$; $A<0, B<0$ and $A<0, 0<B<1$)
$$t_{\infty}=\infty \; ,$$
i.e. the scale factor reaches infinity in infinite time and there are no finite-time singularities. 
\item for $A(1-B) > 0$ there are two cases:
\begin{enumerate}
\item for $\frac{B}{B-1} > 0$ (which corresponds to $A<0, B>1$ and $A>0, B<0$) we obtain
\begin{equation}
\label{eq:tfinite}
t_{\infty}=t_*+\frac{1}{3 | A B|} \sqrt{\frac{3}{8 \pi G \rho_0}}  e^{\frac{(1+w_0)^{1-B}}{2 A (1-B)}} \frac{B}{B-1} (2 A(1-B))^{\frac{B}{B-1}} \Gamma \left( \frac{B}{B-1} \right) \, ,
\end{equation}
i.e. in this regime scale factor becomes infinite in finite time and the expansion ends in finite-time singularity.
\item for $\frac{B}{B-1} < 0$ (corresponding to $A>0, 0 < B <1$)
$$t_{\infty}=\infty \, ,$$
i.e. it takes infinite time for scale factor to become infinite and no finite-time singularities are encountered.
\end{enumerate}
\end{enumerate}

\section{Scalar field representation}

\label{scalar}

Although the description of the dark energy component in terms of a cosmic component is conceptually simple and frequently more analitically tractable compared to other descriptions, it is also of considerable interest to find an equivalent field theory description reproducing the same global expansion of the universe. In this way a correspondence with microscopic, in particular field theory descriptions can be established.

In this section we obtain analytical expressions for the scalar field potential assuming that the only component of the universe is the one defined by (\ref{eq:model}). Although the results obtained based on this assumption are of theoretical interest, they become a useful approximation in the epochs of the universe where the energy density of (\ref{eq:model}) is dominant in the total energy density of the universe. A further assumption is a spatially flat universe, i.e. $k=0$. We proceed separately for the  quintessence-like regime and phantom-like regime. 

\subsection{Quintessence-like regime $(w > -1)$}

For a minimally coupled quintessence scalar field with a Lagrangian ${\cal L}=\frac{1}{2} \dot{\phi}^2 - V(\phi)$ the energy density has the form
\begin{equation}
\label{eq:phirho}
\rho_{\phi}=\frac{\dot{\phi}^2}{2} + V(\phi) \, ,
\end{equation}
 whereas the expression for the pressure is
\begin{equation}
\label{eq:phip}
p_{\phi}=\frac{\dot{\phi}^2}{2} - V(\phi) \, .
\end{equation}

Combining (\ref{eq:phirho}) and (\ref{eq:phip}) with (\ref{eq:cs2ofw}) results in a relation connecting $\phi$ and $w$. In particular

\begin{equation}
\label{eq:dynphiw}
d \phi = \mp \frac{1}{\sqrt{24 \pi G}} \frac{d w}{(c_s^2 -w)\sqrt{1+w}} \, ,
\end{equation}
and for the model (\ref{eq:model})
\begin{equation}
\label{eq:phiofw}
\phi - \phi_0 = \mp \frac{1}{\sqrt{24 \pi G} A}  \frac{1}{-B + \frac{1}{2}} \left((1+w)^{-B + \frac{1}{2}} - (1+w_0)^{-B + \frac{1}{2}} \right) \, .
\end{equation}
It is important to emphasize that in our approach there is no problem with the value $B=1/2$ in (\ref{eq:phiofw}), since it  is not of the form $B=\frac{2 n +1}{2 m +1}$ considered in this paper.The relation (\ref{eq:phiofw}) can be readily inverted which results in
\begin{equation}
\label{eq:wofphi}
w =-1 + \left( (1+w_0)^{-B+\frac{1}{2}} \mp \sqrt{24 \pi G} A\left( -B +\frac{1}{2}\right) (\phi-\phi_0)\right)^{\frac{1}{-B + \frac{1}{2}}} 
\end{equation}

The analytical form of the scalar field potential is then
\begin{eqnarray}
\label{eq:Vofphi}
V(\phi)&=&\frac{\rho_0}{2} \left[ 2 - \left[ (1+w_0)^{-B+\frac{1}{2}} \mp  \sqrt{24 \pi G} A \left(-B +\frac{1}{2} \right) (\phi-\phi_0) \right]^{\frac{1}{-B+\frac{1}{2}}} \right] \nonumber \\
&\times& e^{\frac{1}{A(1-B)} \left[ \left[ (1+w_0)^{-B+\frac{1}{2}} \mp  \sqrt{24 \pi G} A \left(-B +\frac{1}{2} \right) (\phi-\phi_0)\right]^{\frac{-B+1}{-B +\frac{1}{2}}}  - (1+w_0)^{-B+1}\right]}
\end{eqnarray}

\subsection{Phantom-like regime $(w < -1)$}

The minimally coupled phantom scalar filed has a Langrangian ${\cal L}=-\frac{1}{2} \dot{\phi}^2 - V(\phi)$. The energy density is then
\begin{equation}
\label{eq:phirho_ph}
\rho_{\phi}=-\frac{\dot{\phi}^2}{2} + V(\phi) \, ,
\end{equation}
and the expression for pressure is
\begin{equation}
\label{eq:phip_ph}
p_{\phi}=-\frac{\dot{\phi}^2}{2} - V(\phi) \, ,
\end{equation}
Combining (\ref{eq:phirho_ph}) and (\ref{eq:phip_ph}) with (\ref{eq:cs2ofw}) gives a relation between $\phi$ and $w$
\begin{equation}
\label{eq:dynphiw_ph}
d \phi = \pm \frac{1}{\sqrt{24 \pi G}} \frac{d w}{(c_s^2 -w)\sqrt{-1-w}} \, ,
\end{equation}
which, using $(-1)^B=-1$, results in the following expression for the model (\ref{eq:model}):
\begin{equation}
\label{eq:phiofw_ph}
\phi - \phi_0 = \pm \frac{1}{\sqrt{24 \pi G} A}  \frac{1}{-B + \frac{1}{2}} \left((-1-w)^{-B + \frac{1}{2}} - (-1-w_0)^{-B + \frac{1}{2}} \right) \, .
\end{equation}
Inversion of this relation yields an expression  for $w(\phi)$ 
\begin{equation}
\label{eq:wofphi_ph}
w =-1 - \left( (-1-w_0)^{-B+\frac{1}{2}} \pm \sqrt{24 \pi G} A\left( -B +\frac{1}{2}\right) (\phi-\phi_0)\right)^{\frac{1}{-B + \frac{1}{2}}} \, .
\end{equation}
Finally, the potential of the scalar field is
\begin{eqnarray}
\label{eq:Vofphi_ph}
V(\phi)&=&\frac{\rho_0}{2} \left[ 2 + \left[ (-1-w_0)^{-B+\frac{1}{2}} \pm  \sqrt{24 \pi G} A \left(-B +\frac{1}{2} \right) (\phi-\phi_0) \right]^{\frac{1}{-B+\frac{1}{2}}} \right] \nonumber \\
&\times& e^{\frac{1}{A(1-B)} \left[ \left[ (-1-w_0)^{-B+\frac{1}{2}} \pm  \sqrt{24 \pi G} A \left(-B +\frac{1}{2} \right) (\phi-\phi_0)\right]^{\frac{-B+1}{-B +\frac{1}{2}}}  - (-1-w_0)^{-B+1}\right]} \, .
\end{eqnarray}

\section{Discussion}

\label{disc}

The analytical solutions obtained in the preceeding sections allow a detailed discussion which is the focus of this section. We separate our discussion in two subsections. In subsection \ref{disc1} we give a survey of all model parameter regimes with illustrative examples and accompanying plots whereas in subsection \ref{disc2} we concentrate on a specific parameter regime which corresponds to recent fits for dark energy $w(z)$ function to observational data obtained in \cite{NatureAstronomy}.  

\subsection{Parameter regimes}

\label{disc1}

\begin{figure}[!t]
\centering
\includegraphics[scale=0.3]{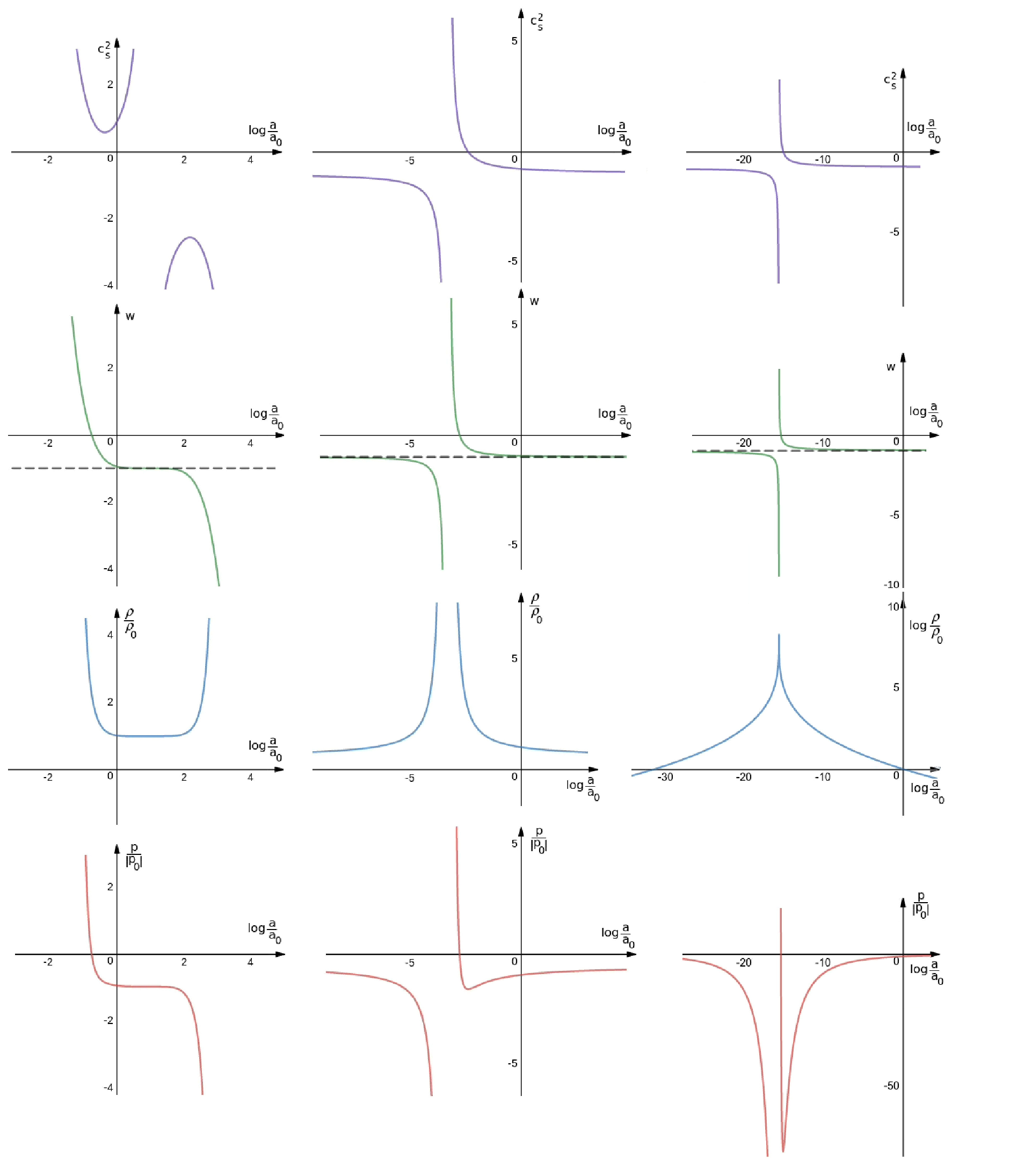}
\caption{Plots of sound speed squared, $c_s^2$ (first row), parameter of  the EoS, $w$ (second row), energy density normalized to its present value, $\rho/\rho_0$ (third row), and pressure normalized to its present absolute value, $p/|p_0|$ (last row), for $A=1$ and $w_0=-0.95$. The left column corresponds to $B=-1/5$, the middle column corresponds to $B=3/5$ and the right column corresponds to $B=7/5$. All plots are log-lin plots except the $\rho(a)$ plot in the third column, which is a log-log plot, to demonstrate the finitness of $\rho$ at $a_*$. 
}
\label{fig1}
\end{figure}

\begin{figure}[!t]
\centering
\includegraphics[scale=0.3]{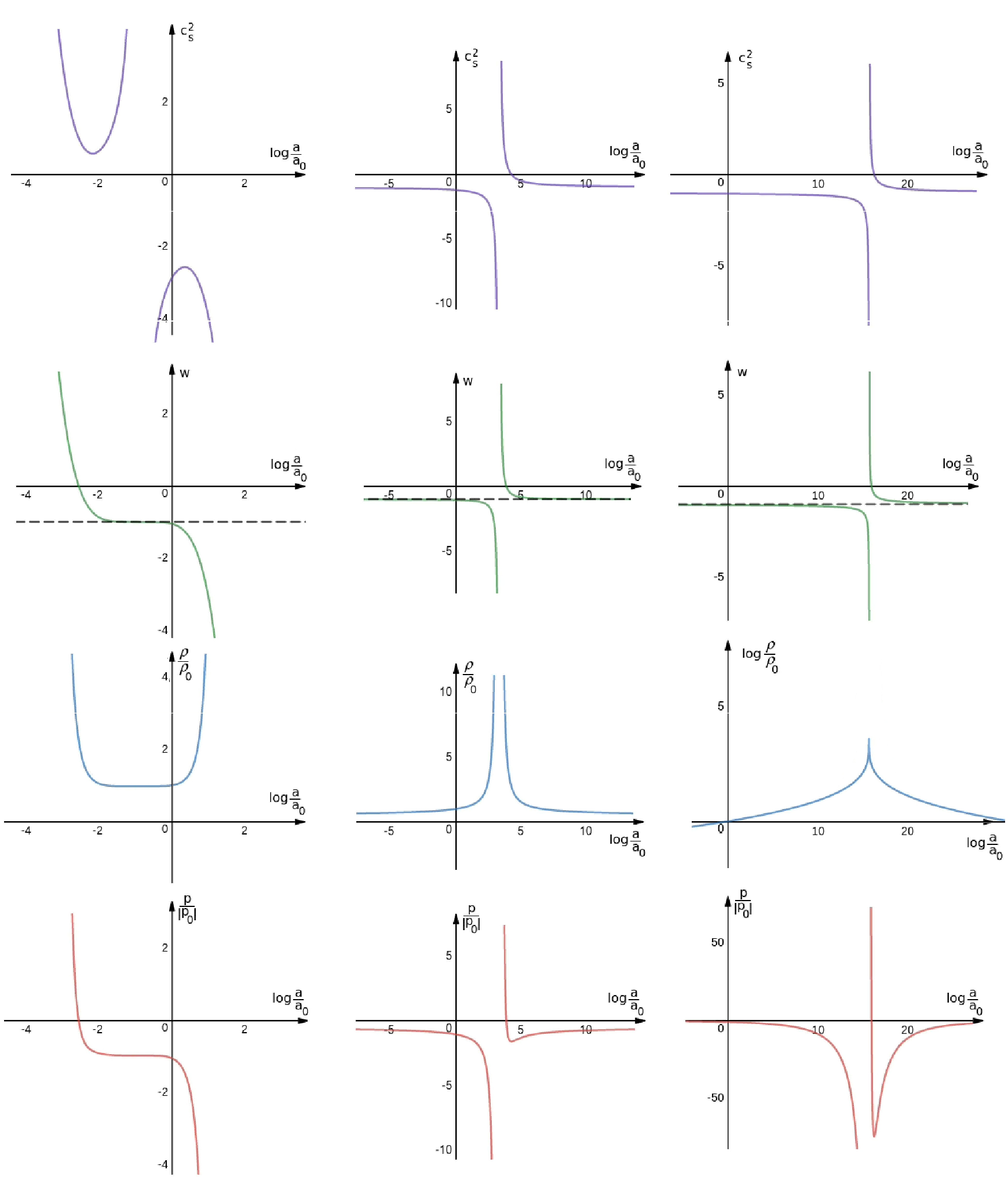}
\caption{Plots of sound speed squared, $c_s^2$ (first row), parameter of  the EoS, $w$ (second row), energy density normalized to its present value, $\rho/\rho_0$ (third row), and pressure normalized to  its present absolute value, $p/|p_0|$  (last row), for $A=1$ and $w_0=-1.05$. The left column corresponds to $B=-1/5$, the middle column corresponds to $B=3/5$ and the right column corresponds to $B=7/5$.  All plots are log-lin plots except the $\rho(a)$ plot in the third column, which is a log-log plot, to demonstrate the finitness of $\rho$ at $a_*$.
}
\label{fig2}
\end{figure}

\begin{figure}[!t]
\centering
\includegraphics[scale=0.3]{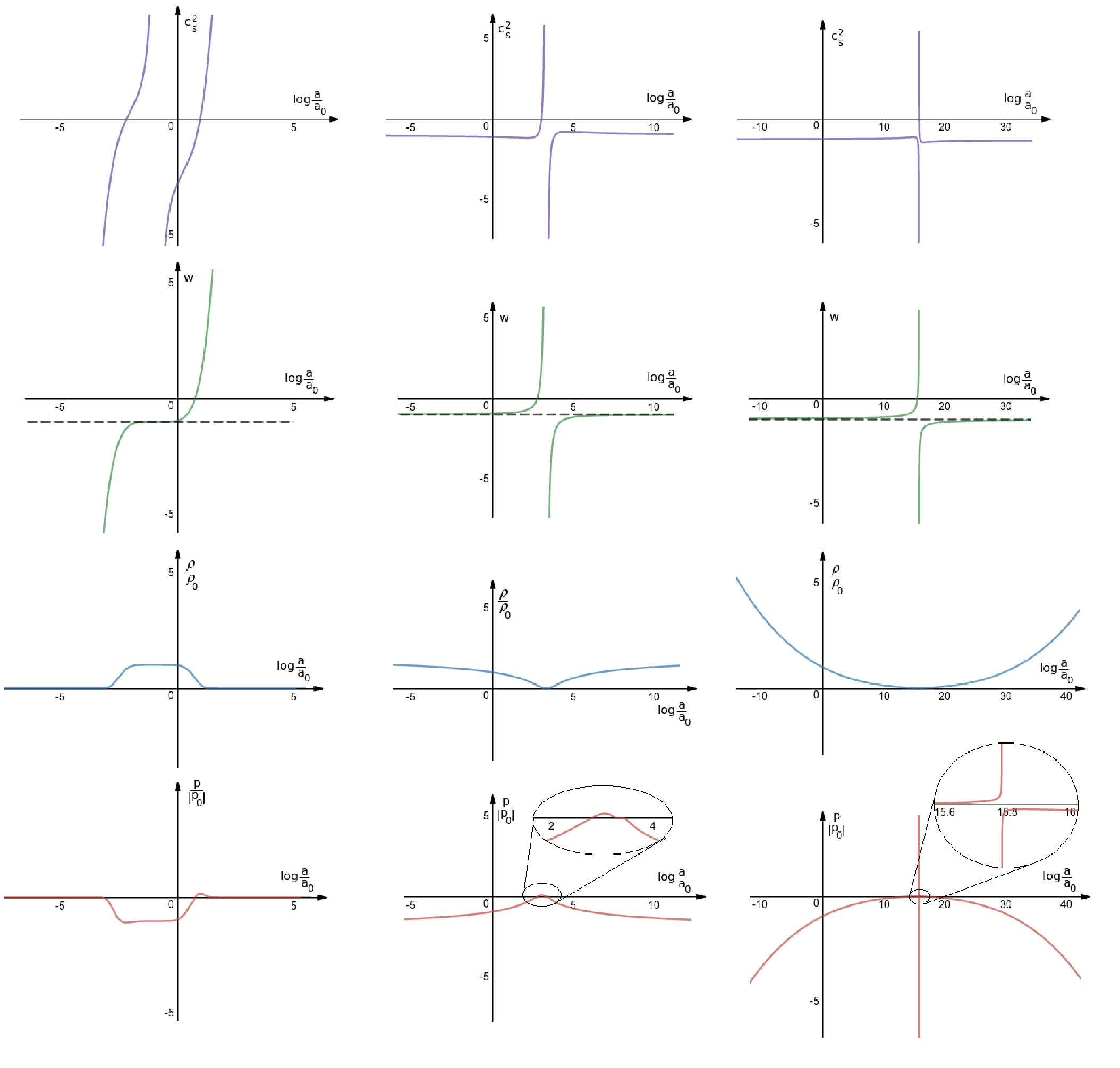}
\caption{Plots of sound speed squared, $c_s^2$ (first row), parameter of  the EoS, $w$ (second row), energy density normalized to its present value, $\rho/\rho_0$ (third row), and pressure normalized to  its present absolute value, $p/|p_0|$  (last row), for $A=-1$ and $w_0=-0.95$. The left column corresponds to $B=-1/5$, the middle column corresponds to $B=3/5$ and the right column corresponds to $B=7/5$.
}
\label{fig3}
\end{figure}

\begin{figure}[!t]
\centering
\includegraphics[scale=0.3]{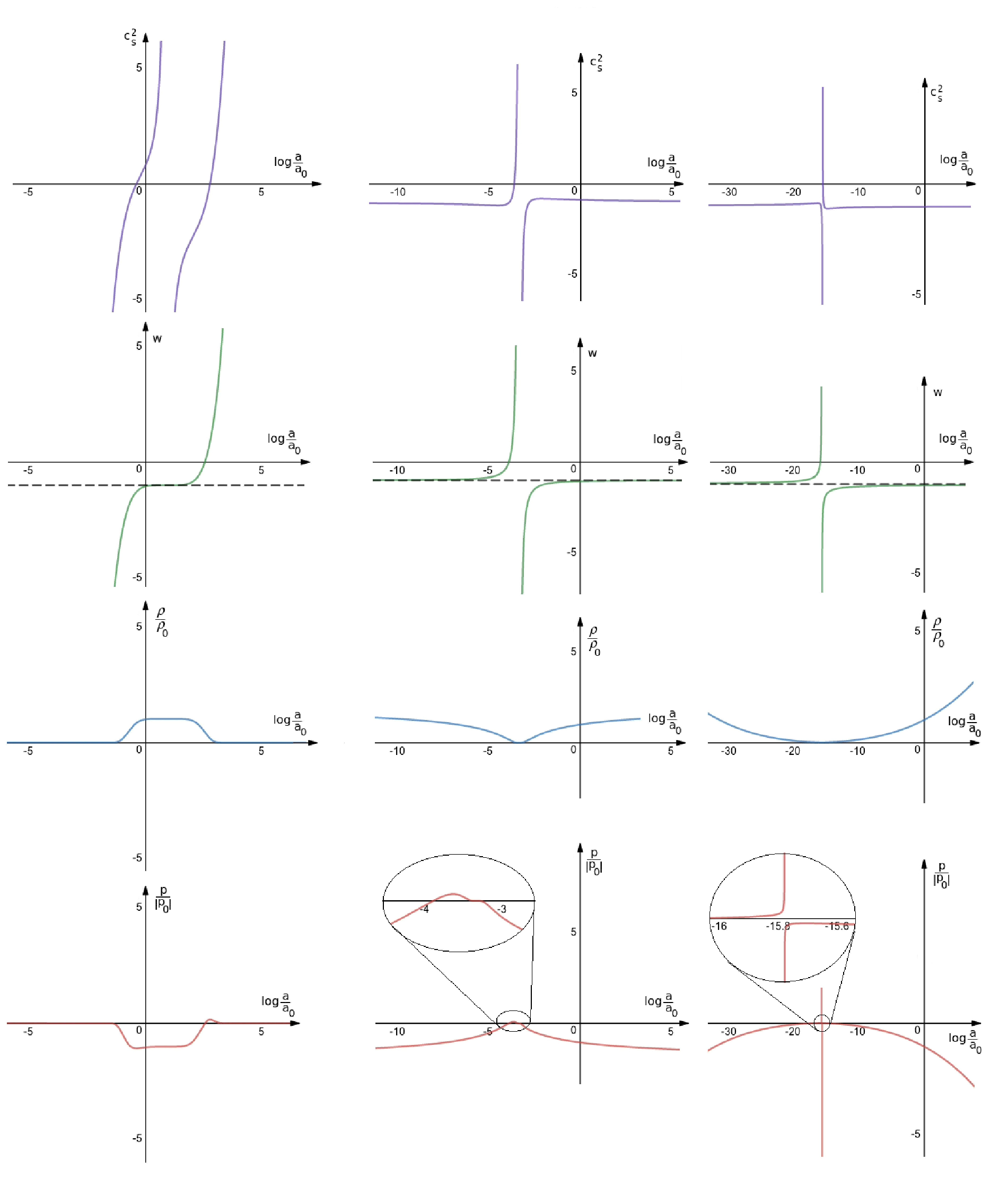}
\caption{Plots of sound speed squared, $c_s^2$ (first row), parameter of  the EoS, $w$ (second row), energy density normalized to its present value, $\rho/\rho_0$ (third row), and pressure normalized to  its present absolute value, $p/|p_0|$  (last row), for $A=-1$ and $w_0=-1.05$. The left column corresponds to $B=-1/5$, the middle column corresponds to $B=3/5$ and the right column corresponds to $B=7/5$.
}
\label{fig4}
\end{figure}

The inspection of solutions (\ref{eq:wofa}), (\ref{eq:rhoofa}) and (\ref{eq:tofa}) indicates twelve regimes for the model parameters $A$, $B$ and $w_0$. In particular, we could expect different dynamics for two intervals of $w_0$ ($w_0<-1$ and $w_0>-1$), two intervals of $A$ ($A<0$ and $A>0$) and three intervals of $B$ ($B<0$, $0<B<1$ and $B>1$). The behavior of the model for parameter values representative for these twelve regimes is presented in Figures  \ref{fig1} to \ref{fig4}. Apart from the asimptotic values $a \rightarrow 0$ and $a \rightarrow \infty$, a specific value of  interest is the scale factor value for which the function $(1+w_0)^{-B} + 3 A B \ln \frac{a}{a_0}$ vanishes i.e.
\begin{equation}
\label{eq:astar}
a_{*}=a_0 e^{-\frac{(1+w_0)^{-B}}{3 A B}} \, .
\end{equation}
The impact of the parameter $w_0$ is essentially reflected in the shifts of plots of  $c_s(a)$, $w(a)$, $\rho(a)$ and $p(a)$ along the $a$ axis. Therefore, in the remainder of this section we focus on the anaysis of the impact of parameters $A$ and $B$. The regimes of interest are the following:  
\begin{itemize}
\item $A>0$, $B<0$. Representative examples of this regime are depicted in the left column of Figures \ref{fig1} and \ref{fig2}. The analysis of (\ref{eq:wofa}) reveals a CC boundary crosing from quintessence-like to the phantom-like regime with $\lim_{a \rightarrow 0} w =  + \infty$, $\lim_{a \rightarrow \infty} w =  - \infty$ and $\lim_{a \rightarrow a_*} w =  -1$. The scaling of  energy density presented in (\ref{eq:rhoofa}) shows a decreasing function which levels around present time and then grows with the expansion of the universe. In particular, $\lim_{a \rightarrow 0} \rho =  + \infty$, $\lim_{a \rightarrow \infty} \rho =  + \infty$ and $\lim_{a \rightarrow a_*} \rho =  \rho_0 e^{-\frac{(1+w_0)^{-B}}{A(1-B)}}$. Correspondingly, pressure transits from positive to negative values with the expansion of the universe. An especially interesting feature of model dynamics in this regime is a plateau where $w \simeq -1$, $\rho \simeq const$ and $p \simeq const$ in which the model closely mimics the cosmological constant for a (relatively) broad interval of scale factor. A similar phase of CC-like behavior for scalar field models with exponential potentials was described in \cite{Russo}. Some other dynamical DE models in which dark energy is very close to cosmological constant, capable of explaining many features of observational data, comprise Running vacuum models \cite{RVM1,RVM2,RVM3}. The analysis of section \ref{scalefactor} shows that in this regime of our model $t_{\infty}$ is finite i.e. the dynamics of $a(t)$ ends in Big Rip singularity \cite{cald1,cald2} in finite time.

\item $A>0$, $0<B<1$. This regime is presented in middle columns of Figures \ref{fig1} and \ref{fig2}. The dynamics of $w$ with scale factor reveals a singularity in $w$ and no CC boundary crossing. The details of dynamics include $\lim_{a \rightarrow 0} w =  -1$, $\lim_{a \rightarrow \infty} w =  - 1$, $\lim_{a \rightarrow a_{*}^{-}} w =  -\infty$ and $\lim_{a \rightarrow a_{*}^{+}} w =  +\infty$. The energy density dynamics exhibits finite asymptotic values, i.e. $\lim_{a \rightarrow 0} \rho = \rho_0 e^{-\frac{(1+w_0)^{-B}}{A(1-B)}} $ and $\lim_{a \rightarrow \infty} \rho =  \rho_0 e^{-\frac{(1+w_0)^{-B}}{A(1-B)}}$, and a singularity at $a_*$, i.e. $\lim_{a \rightarrow a_{*}^{-}} \rho = +\infty$ and $\lim_{a \rightarrow a_{*}^{+}} \rho = +\infty$. It is interesting to notice that in this regime the model dynamics resembles the cosmological constant everywhere outside the vicinity of the singularity. The scale factor dynamics exhibits no finite time singularities, as indicated in section \ref{scalefactor}. Since in (\ref{eq:tinfty}) as a limit of integration we use $a_*$ at which $\rho$ diverges, it is reasonable to ask if this fact affects the calculation of $t_{\infty}$. However, direct integration of (\ref{eq:tofa}) in a short interval around $a_*$ (with $\tilde{a} < a_* < a$, i.e. $\tilde{t} < t_* < t$) shows that corresponding time interval $t - \tilde{t}$ is finite and, therefore, singular behavior of $\rho$ at $a_*$ cannot interfere with the result that $t_{\infty}=\infty$. 

\item $A>0$, $B>1$. This regime corresponds to the right column of Figures \ref{fig1} and \ref{fig2}. The function $w(a)$ is singular at $a_*$ and asymptotically tends to -1. In particular, $\lim_{a \rightarrow 0} w =  -1$, $\lim_{a \rightarrow \infty} w =  - 1$, $\lim_{a \rightarrow a_{*}^{-}} w =  -\infty$ and $\lim_{a \rightarrow a_{*}^{+}} w =  +\infty$. The energy density asymptotically vanishes, i.e. $\lim_{a \rightarrow 0} \rho = 0 $ and $\lim_{a \rightarrow \infty} \rho = 0$, reaching a maximum at $a_*$, i.e. $\lim_{a \rightarrow a_{*}} \rho =  \rho_0 e^{-\frac{(1+w_0)^{-B}}{A(1-B)}}$. It is important to stress that the dynamics of $\rho(a)$ is nonsingular, despite the existence of singularity in $w(a)$. Naturally, pressure exhibits singularity at $a_*$. The dynamics of $\rho(a)$ corresponds to the concept of transient dark energy component introduced in section \ref{intro}.  The analysis of section \ref{scalefactor} shows $t_{\infty} \rightarrow \infty$, i.e. the absence of future singularities. 

\item $A<0$, $B<0$. Examples of this regime are given in left columns of Figures \ref{fig3} and \ref{fig4}. The dynamics of $w(a)$ exhibits the CC boundary crossing with transition from the phantom-like regime to the quintessence-like regime with $\lim_{a \rightarrow 0} w =  - \infty$, $\lim_{a \rightarrow \infty} w =  + \infty$ and $\lim_{a \rightarrow a_*} w =  -1$. The energy density has vanishing asymptotic values, i.e. $\lim_{a \rightarrow 0} \rho =  0$ and $\lim_{a \rightarrow \infty} \rho =  0$, with $\lim_{a \rightarrow a_*} \rho =  \rho_0 e^{-\frac{(1+w_0)^{-B}}{A(1-B)}}$. As the universe expands, the presure, starting asymptotically from zero, first becomes increasingly negative, then stabilizes and finally transits to positive values, reaches a maximal positive value and asymptotically returns to zero. A particullarly interesting feature of this regime is a (relatively) broad range of scale factor  in which the model dynamics closely resembles the cosmological constant. These features reveal that our DE model in this regime has the properties of transient dark energy. The scale factor dynamics reveals no future finite time singularities. This regime of model (\ref{eq:model}) will be further discussed in the following subsection in terms of recent fits of $w(z)$ reported in \cite{NatureAstronomy}.

\item $A<0$, $0<B<1$. Examples coresponding to this regime are presented in middle columns of Figures \ref{fig3} and \ref{fig4}. The function $w(a)$ has asymptotic values of -1, i.e. $\lim_{a \rightarrow 0} w =  -1$ and $\lim_{a \rightarrow \infty} w =  - 1$, and a singularity at $a_*$, i.e. $\lim_{a \rightarrow a_{*}^{-}} w =  +\infty$ and $\lim_{a \rightarrow a_{*}^{+}} w =  -\infty$. The energy density dynamics has finite asymptotic values, $\lim_{a \rightarrow 0} \rho = \rho_0 e^{-\frac{(1+w_0)^{-B}}{A(1-B)}} $ and $\lim_{a \rightarrow \infty} \rho =  \rho_0 e^{-\frac{(1+w_0)^{-B}}{A(1-B)}}$, whereas it vanishes at $a_*$, i.e. $\lim_{a \rightarrow a_{*}} \rho = 0$. It is interesting to see that pressure function does not exhibit singularities and the singularity in $w$ is a consequence of vanishing $\rho$. The scale factor reaches infinity in infinite time, i.e. $t_{\infty}=\infty$. It is also interesting to consider the dynamics of $\rho$ for $a>a_*$ as an instance of a Pseudo-Rip proposed in \cite{PR}.

\item $A<0$, $B>1$. This regime is presented in the right column of Figures \ref{fig3} and \ref{fig4}. The dynamics of $w(a)$ is characterized by asymptotic values of $-1$, i.e. $\lim_{a \rightarrow 0} w =  -1$ and $\lim_{a \rightarrow \infty} w =  - 1$, with a singularity at $a_*$, i.e. $\lim_{a \rightarrow a_{*}^{-}} w =  +\infty$ and $\lim_{a \rightarrow a_{*}^{+}} w =  -\infty$. The energy density grows without limit asymptotically, $\lim_{a \rightarrow 0} \rho = +\infty $ and $\lim_{a \rightarrow \infty} \rho = +\infty$ with $\lim_{a \rightarrow a_{*}} \rho =  \rho_0 e^{-\frac{(1+w_0)^{-B}}{A(1-B)}}$. In this regime, presure is also singular at $a_*$.  The analyisis of section \ref{scalefactor} reveals a finite value of $t_{\infty}$, i.e. a Big Rip singularity in finite time. 

\end{itemize}

\begin{figure}[!t]
\centering
\includegraphics[scale=0.3]{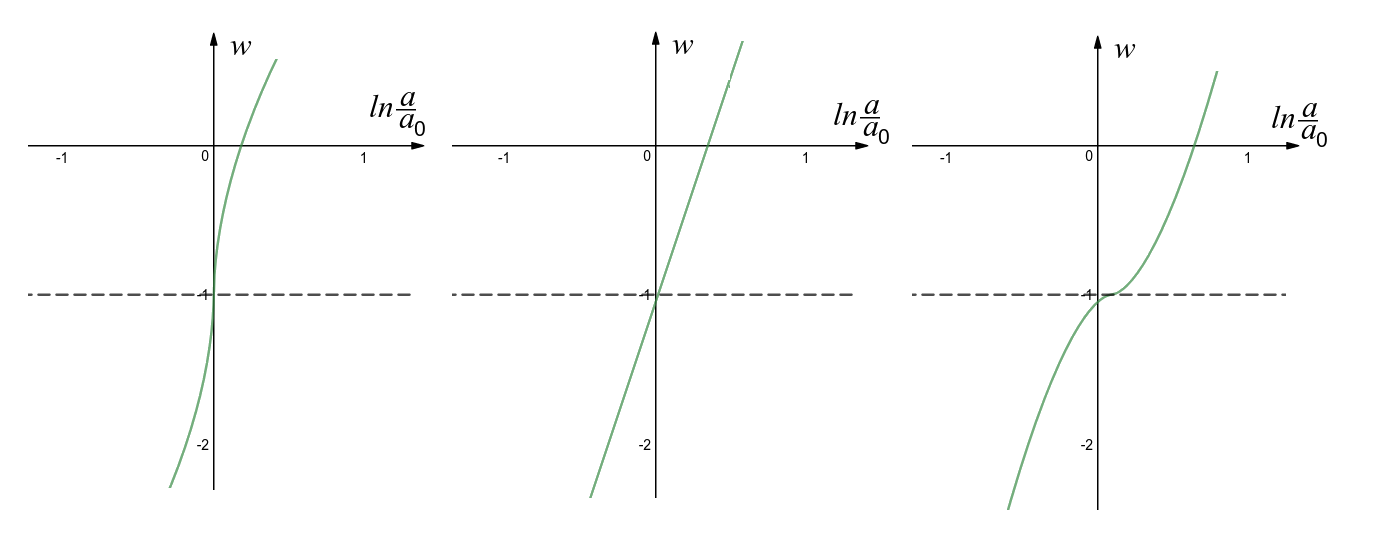}
\caption{Plots of parameter of the EoS $w$ as a function of the scale factor for $A<0$ and for three values of the parameter $B$  in the $B<0$ regime: $B=-9/5$ (left), $B=-1$ (middle) and $B=-3/5$ (right).
}
\label{fig:Bminus1}
\end{figure}

In the region $B<0$, Eq. (\ref{eq:wofaexplicit}) provides additional insight into the properties of the $w(a)$ function at the CC boundary crossing point. This expression reveals $B=-1$ as another point of interest. At the crossing point where $1+w=0$ the derivative $ a \frac{d \, w}{d \, a}$ equals
\begin{itemize}
\item $0$ for $B>-1$ , 
\item $-3 A$ for $B=-1$,
\item $\mathrm{sign}(-3A) \cdot \infty$ for $B<-1$.
\end{itemize}
The dependence of $w(a)$ on the parameter $B$ in the region $B<0$ is presented in Fig. \ref{fig:Bminus1}.

A few remarks on the behavior of $c_s^2$ function, given in (\ref{eq:cs2ofa}), are in order. The plots in the first row of Figures \ref{fig1}-\ref{fig4} depict the dependence of this quantity on the scale factor. A common property of all these plots is an existence of a singularity in $c_s^2$, with possible associated issues with causality. It is interesting that the singularity in $c_s^2$ appears both in cases when $w(a)$ exhibits singular behavior (such as for regimes with $B>0$) and when $w(a)$ is nonsingular (such as for regimes with $B<0$). In the former case ($B>0$), $c_s^2$ ``inherits'' the singularity appearing in $w(a)$, whereas in the latter case ($B<0$), the singularity in $c_s^2$ appears when otherwise nonsingular $w(a)$ crosses $w=-1$ boundary. Although these conclusions strictly apply to model (\ref{eq:model}) studied in this paper, an intriguing question for future work is if such a  singular behavior of $c_s^2$ is a generic feature of all models in which CC boundary crossing is described in terms of a single cosmic component.  Furthermore, $c_s^2$ acquires negative values in some intervals of the scale factor, which lead to instabilities of perturbations at small scales for perfect fluid models. These facts indicate that $c_s^2$ should not be interpreted as the physical speed of sound and the entire approach serves as an effective way of modeling $d p/d \rho$ as a function of $p/\rho$. 

Finally, a natural question arises from Eq. (\ref{eq:conserv}). At first sight this equation indicates that $w=-1$ is the fixed point of evolution of the single component  and that it cannot be crossed. The same conclusion might seem to follow from the existence of the $1+w$  term in the denominator of the left hand side of Eq. (\ref{eq:cs2ofw}), which at first sight leads to a nonintegrable singularity at $w=-1$. The existence of solutions presented in this paper for $B<0$ clearly shows that the conclusion of $w=-1$ being the fixed point is not generally true and that some cancellations must be at work for the $B<0$ regime of our model. Indeed, these cancellations become clear when inserting (\ref{eq:model}) for $B<0$ in (\ref{eq:cs2ofw}). Now the LHS of (\ref{eq:cs2ofw}) has an integrable singularity or no singularity at all in $w=-1$. The $c_s^2(w)$ modeling approach has a clear advantage that it is easy to choose a $c_s^2(w)$ function for which the aforementioned cancellation can be achieved.

\subsection{Empirical dark energy $w(z)$ functions}

\label{disc2}

In a recent analysis, published in \cite{NatureAstronomy}, the problem of reducing tension in estimation of cosmological parameters from observational sets corresponding to different cosmic phenomena was addressed. In particular, an attempt to reconcile cosmic parameter inference from these different datasets was  made by assuming a general redshift dependence of dark energy EoS parameter, $w(z)$. In practice, the range of redshifts covered by the analysis is divided into subintervals and $w$ is assumed to be constant within each of these subintervals. The values of $w$ in subintervals of redshift then become parameters of the model.    

Such a statistical analysis yields a wiggly $w(z)$ function which intersects $w=-1$ line several times, as presented in Fig. 2 of \cite{NatureAstronomy}. This very fact makes the understanding of the CC boundary crossing phenomenon a very interesting topic by itself. However, in terms of statistical significance of the deviation of $w(z)$ from $-1$, we could divide the interval of redshift studied in \cite{NatureAstronomy} to subinterval $0 < z \le 1.2$ in which $w \approx -1$ and subinterval $z > 1.2$ in which $w(z)$ turns to values significantly more negative than $-1$. 

\begin{figure}[!t]
\centering
\includegraphics[scale=0.3]{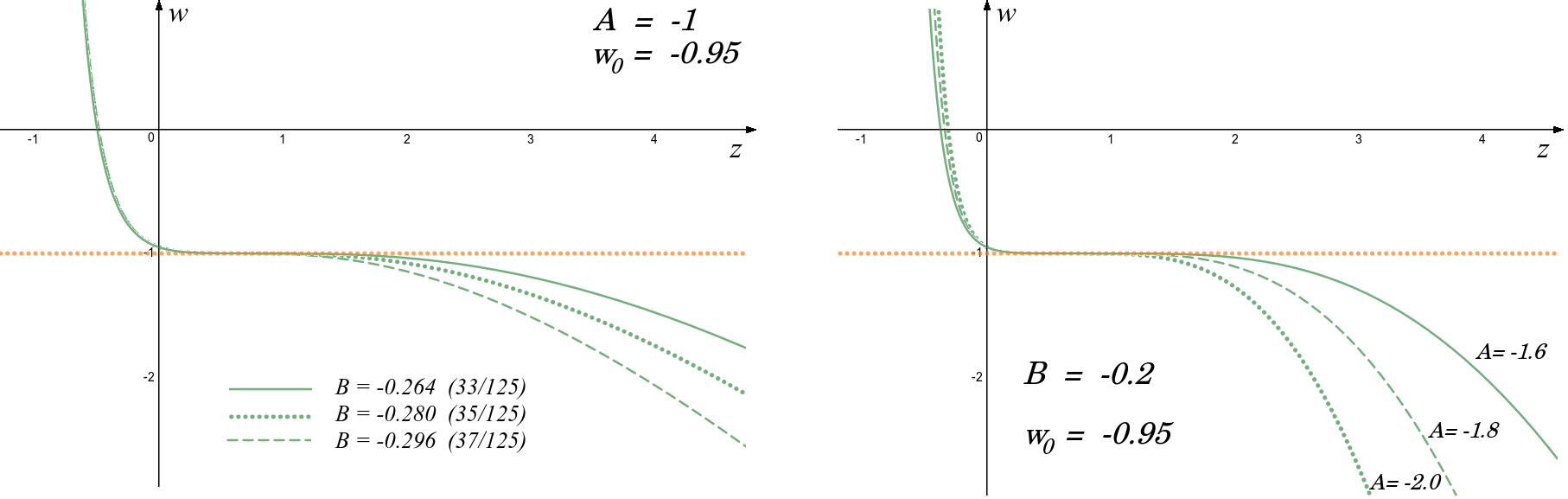}
\caption{The redshift dependence of the EoS parameter, $w(z)$ in the regime $A<0, B<0$ reproducing general features of $w(z)$ fit function from \cite{NatureAstronomy}. The left plot shows the influence of variation in parameter $B$, whereas the right plot presents the influence of variation in parameter $A$.
}
\label{fig5}
\end{figure}

It is important to stress that such a behavior of $w(z)$ (very close to -1 for an interval of $z$ close to 0, then turning to phantom-like values for larger $z$ values) is characteristic of our model for parameters $A<0, B<0$, as depicted in left column of Figures \ref{fig3} and \ref{fig4}. A more detailed dynamics of $w$ in this parameter regime of our model is presented in 
Fig. \ref{fig5}. Here parameter values are chosen for $w(z)$ to resemble the general features of the $w(z)$ fit function from \cite{NatureAstronomy}. Precise parameter values should be obained from the full statistical analysis which is out of scope of this paper.

It is interesting to observe that the regime $A<0, B<0$ of the model corresponds to transient dark energy and it is an intriguing to ask if the form of $w(z)$ fit function from \cite{NatureAstronomy} is an indication of transient dark energy.

\section{Conclusions}

\label{concl}

In this paper a novel approach to the modeling of dark energy dynamics was tested at several levels. In particular, building on results of \cite{PRD2013}, it is shown that modeling of speed of sound as a function of the EoS parameter, i.e. $c_s^2=c_s^2(w)$ is useful as a framework for studying particular properties of dark energy. For a concrete model (\ref{eq:model}) analytical solutions for dark energy dynamics are obtained. These solutions shed light on the phenomenon of cosmological constant boundary crossing and the related phenomenon of transient dark energy. One of important results of the paper is a description of the phenomenon of CC boundary crossing in a simple analitically tractable model.  It is further shown that for the particular studied model the $c_s^2(w)$ is singular in all parameter regimes and it also exhibits intervals where it is negative. This lends support to an effective view on the $c_s^2(w)$ modeling approach. i.e. interpreting it as modelling of $dp/d\rho$ as a function of $p/\rho$. A particular regime of model (\ref{eq:model}), $A<0, B<0$, corresponding to transient dark energy, is shown to produce $w(z)$ function that describes general features of the recently reported fits to observational data for the dark energy $w(z)$ function. This fact opens up an interesting possibility that the universe that we live in is actually endowed with a transient dark energy. The validation of this possibility deserves future research efforts within the $c_s^2(w)$ formalism.

\end{document}